\documentstyle[12pt]{article}
\newcommand{\abs}[1]{\left| #1 \right|}
\newcommand{\VEV}[1]{\left<#1\right>}
\newcommand{\order}[1]{{\cal O}\!\left(#1\right)}
\begin{document}
\begin{titlepage}
\begin{flushright}
 KUNS-1863\\
 NIIG-DP-03-04
\end{flushright}

\begin{center}
\vspace*{5mm}
  
{\LARGE\bf
Anomalous $\mathbf{U(1)}$ $\mathbf{D}$-term Contribution\\[7pt]
in Type~I String Models}
\vspace{12mm}

{\large
Tetsutaro~Higaki,\footnote{
E-mail address: tetsu@gauge.scphys.kyoto-u.ac.jp}
Yoshiharu~Kawamura,\footnote{
E-mail address: haru@azusa.shinshu-u.ac.jp}
Tatsuo~Kobayashi\footnote{
E-mail address: kobayash@gauge.scphys.kyoto-u.ac.jp}\\[5pt]
and~~Hiroaki~Nakano\footnote{
E-mail address: nakano@muse.sc.niigata-u.ac.jp}
}
\vspace{6mm}

{\it $^{1,3}$Department of Physics, Kyoto University,
Kyoto 606-8502, Japan}\\[1mm]
{\it $^2$Department of Physics, Shinshu University,
Matsumoto 390-8621, Japan}\\[1mm]
{\it $^4$Department of Physics, Niigata University,
Niigata 950-2181, Japan}

\vspace*{15mm}

\begin{abstract}
We study the $D$-term contribution 
for anomalous $U(1)$ symmetries in type~I string models 
and derive general formula for the $D$-term contribution, 
assuming that the dominant source of SUSY breaking is given by 
$F$-terms of the dilaton, (overall) moduli or twisted moduli fields.
On the basis of the formula,
we also point out that there are several different features 
from the case in heterotic string models.
The differences originate from 
the different forms of K\"ahler potential 
between twisted moduli fields in type~I string models
and the dilaton field in heterotic string models.
\end{abstract}

\end{center}
\end{titlepage}

\section{Introduction}
\label{sec:introduction}

Superstring theory is a promising candidate 
for unified theory including gravity.
One of important features is that 4-dimensional (4D) string models 
have several moduli fields including the dilaton 
field.
Their vacuum expectation values (VEVs) determine 
couplings of 4D effective theory, e.g. gauge couplings, 
Yukawa couplings and Fayet-Iliopoulos (FI) coefficients.
These moduli fields have perturbatively a flat potential.
Non-perturbative effects are expected to stabilize these 
moduli.
Such non-perturbative effects may also break 
supersymmetry (SUSY) at the same time.
If SUSY is broken, SUSY breaking terms, 
e.g. gaugino masses and soft scalar masses, are induced.
The pattern of SUSY breaking terms depends 
on couplings of gauge and matter fields to moduli fields.
These s-spectra would be measured in near future. 
Thus, it is very important to study SUSY breaking terms 
in 4D string models.

Actually, such analyses have been done extensively  
both in heterotic models \cite{CCM,string} and 
in type~I models \cite{IIB}.
{}For example, the dilaton-dominant SUSY breaking in 4D 
heterotic models 
has high predictability, when we consider the scalar 
potential only due to $F$-terms.
That leads to the universal relation, 
$M^\alpha_{1/2} = -A_{IJK} = \sqrt 3 m_{3/2}$ 
and $m^2_I = |m_{3/2}|^2$, 
where $m_{3/2}$ is the gravitino mass, $M^\alpha_{1/2}$ is 
the gaugino mass, $A_{IJK}$ is the $A$-term and $m_I$ is 
soft scalar masses,  while SUSY breaking due to other 
sources leads to non-universal relations.
The universal spectrum of sfermion masses 
is favorable from the viewpoint of flavor changing 
neutral current (FCNC) constraints.
On the other hand, the high predictability may face 
problems.
{}For example, this pattern of SUSY breaking terms 
easily leads to color and/or charge breaking (CCB) 
or the  unbounded from below (UFB) 
direction \cite{Casas:1996wj}.\footnote{
This problem would not be serious, if the age of 
the Universe is not long enough to reach the CCB minimum.}
Similarly, SUSY breaking terms have been studied 
in type~I models when we consider the scalar potential 
only due to $F$-terms \cite{IIB,King:2001zi}.

Most of 4D string models for both heterotic models 
and type~I models have anomalous $U(1)$ 
symmetries \cite{DSW,KN,typeI}.
Many 4D type~I models have been built 
e.g. through the type~IIB orientifold construction.
The anomaly is cancelled by the Green-Schwarz mechanism, 
where certain fields transform non-linearly.
Such role is played by the dilaton field in heterotic 
models and twisted moduli fields in type~I models, respectively.
Then these fields generate FI terms, whose magnitudes 
are determined by VEVs of the dilaton field and 
twisted moduli fields.
Other chiral matter fields develop their VEVs along 
the almost $D$-flat direction 
($D$-flat direction in the SUSY limit) 
and $U(1)$ symmetries are broken.
As a phenomenological application of anomalous $U(1)$ 
symmetry, it can be used as a flavor symmetry 
for the Froggatt-Nielsen mechanism \cite{FN,IR}.
If one can assign $U(1)$ charges suitably to quarks and 
leptons, realistic Yukawa matrices can be derived.

In general,  there appears an additional contribution to 
soft SUSY breaking scalar masses
called the $\lq\lq$$D$-term contribution" after gauge symmetries 
are broken down \cite{D-term,KMY}.
This contribution has a linear dependence on the VEV of 
$D$-component
and it is proportional to the charge of broken symmetry.
These features are different from those in the 
contribution from $F$-component,
which has the quadratic form of the VEVs of $F$-component 
and it does not depend on the charge of broken symmetry 
explicitly.
A magnitude of $D$-term condensation has been studied
in grand unified theories \cite{KMY,JKY}.

Since most of 4D string models have 
an anomalous $U(1)$ symmetry, its breaking, in general, 
induces a $D$-term contribution to scalar masses.
{}For 4D heterotic models, the $D$-term contribution 
has been examined \cite{DA,KK,DPS,K}.
In particular, in Ref.~\cite{KK} 
it is taken into account that 
the FI term is dilaton-dependent.
As a result, even in the dilaton-dominant SUSY breaking 
the $D$-term contribution induces non-universal 
scalar masses, and the additional terms are 
proportional to $U(1)$ charges.
That has phenomenologically important implications.
{}For example, the CCB and UFB constraints can be 
relaxed \cite{CCB}.
As another aspect, these $D$-term contributions 
have an important implication 
for the Froggatt-Nielsen mechanism.
In order to derive realistically hierarchical Yukawa 
matrices, one has to assign different $U(1)$ charges 
for different families.
In this case, the $D$-term contribution proportional 
to $U(1)$ charges leads to non-universal sfermion 
masses, which are dangerous from the viewpoint of FCNC constraints.


In this way, it is an important subject to study
a magnitude of $D$-term condensation for each model.
In this paper, we study the $D$-term contribution for 
anomalous $U(1)$ symmetries
in 4D type~I models
and point out that there are several different features from
the case in heterotic models.
Such difference comes from the fact that in type~I models 
the twisted moduli fields play a role in the Green-Schwarz (GS) 
anomaly cancellation mechanism, that is, 
the FI term depends on the twisted moduli fields.
Their $F$-components can contribute to SUSY breaking.\footnote{
See Ref.~\cite{Abel:2000tf,Ciesielski:2002fs,Higaki:2003zk} 
for SUSY breaking scenario by the $F$-term of the twisted moduli fields.}
Their K\"ahler potential is expected to be different from 
that of the dilaton field.
Furthermore, unlike the dilaton VEV in 
heterotic models,
the VEVs of twisted moduli fields 
can be taken freely.

This paper is organized as follows. 
In the next section, we explain the $D$-term contribution 
to soft SUSY breaking scalar masses and the general formula 
for the VEV of the $D$-auxiliary fields.
After reviewing the $D$-term contribution for the anomalous 
$U(1)$ symmetry based on heterotic models in section~\ref{sec:het}, 
we study the $D$-term contribution for anomalous $U(1)$ symmetries
in the framework of type~I models in section~\ref{sec:typeI}.
In section~\ref{sec:implications}, we discuss 
phenomenological implications of $D$-term contributions.
Section~\ref{sec:conclusion} is devoted to conclusion.

\section{$D$-term contribution}
\label{sec:formula}

We explain the $D$-term contribution to soft SUSY breaking 
scalar masses
based on supergravity theory (SUGRA) \cite{JKY}.
The matter sector in SUGRA  is specified by two functions, 
the total K\"ahler
potential $G(\phi^I, \bar{\phi}^{\bar{I}})$ and the gauge 
kinetic function
$f_{\alpha \beta}(\phi^I)$ with $\alpha$, $\beta$  being 
indices of 
the adjoint representations of the gauge groups.   
The former is a sum of the K\"ahler potential 
$K(\phi^I, \bar{\phi}^{\bar{I}})$ 
and the logarithm of the superpotential $W(\phi^I)$
\begin{eqnarray}
G(\phi^I, \bar{\phi}^{\bar{I}})=K(\phi^I, \bar{\phi}^{\bar{I}}) 
+ M^{2}\ln \frac{|W (\phi^I) |^2}{M^{6}} \ ,
\label{total-Kahler}
\end{eqnarray}
where $M$ is the gravitational scale defined by use of the 
Planck mass $M_{Pl}$
such as $M \equiv M_{Pl}/\sqrt{8\pi}$.
We have denoted scalar fields in the chiral multiplets by 
$\phi^I$
and their complex conjugate by $\bar{\phi}^{\bar{I}}$.
The real part of gauge kinetic function 
${\mbox {Re}}f_{\alpha\beta}$ 
is related to the gauge coupling constants $g_{\alpha}$ 
as follows,
\begin{eqnarray}
  \langle {\mbox{Re}}f_{\alpha\beta} \rangle = 
\frac{1}{g_\alpha^2}\,\delta_{\alpha\beta} \ .
\label{gauge-kinetic}
\end{eqnarray}
The scalar potential is given by
\begin{eqnarray}
V = M^{2}e^{G/M^{2}} 
    \left(G_I (G^{-1})^{I\bar{J}} G_{\bar{J}}-3M^{2}\right)
  + \frac{1}{2} \left(\mbox{Re} f^{-1}\right)_{\alpha \beta}
    G_{I\,}(T^\alpha\phi)^I G_J(T^\beta \phi)^J \ ,
\label{V-SUGRA}
\end{eqnarray}
where
$G_I=\partial G/\partial \phi^I$, 
$G_{\bar{I}}=\partial G/\partial \bar{\phi}^{\bar{I}}$ etc,  
$({\mbox {Re}} f^{-1})_{\alpha \beta}$ and  
$(G^{-1})^{I\bar{J}}$ are the inverse matrices of 
$\mbox{Re} f_{\alpha \beta}$ and  
$G_{I\bar{J}}$, respectively, e.g., 
$(G^{-1})^{I\bar{J}'}G_{\bar{J}'I'} = 
\delta^I_{I'}$,
$T^\alpha$ are gauge transformation generators,
and a summation over $\alpha$,... and  $I$,... is understood.  
The $( T^\alpha \phi)^I$ are variations under gauge 
transformations
up to infinitesimal parameters and they become constants for 
non-linear
realizations.
The $F$-auxiliary fields and 
the $D$-auxiliary fields are given by
\begin{eqnarray}
{} F^I
 =  Me^{G/2M^{2}} (G^{-1})^{I\bar{J}} G_{\bar{J}} \ , \qquad
D^{\alpha}
 = (\mbox{Re} f^{-1})_{\alpha\beta} G_I ( T^\beta \phi)^I \ ,
\label{D}
\end{eqnarray}
respectively. In terms of $F^I$ and $D^\alpha$, 
the scalar potential takes the form
\begin{eqnarray}
V = V_F + V_D \equiv
    \left(\bar{F}^{\bar{J}} K_{\bar{J}I} F^I - 3M^{4} e^{G/M^{2}}
    \right)
  + \frac{1}{2}\,\mbox{Re} f_{\alpha \beta} D^{\alpha} D^{\beta} \ .
\label{V}
\end{eqnarray}

By taking the flat limit of $V$, we obtain the soft SUSY 
breaking terms for scalar fields.
Here we are interested in the scalar mass terms
\begin{eqnarray}
V_{{\rm soft}}
 &=& (m^2_F)_{I \bar{J}\,} \phi^I \bar{\phi}^{\bar{J}} 
    +(m^2_D)_{I \bar{J}\,} \phi^I \bar{\phi}^{\bar{J}} + \cdots \ ,
\label{m}\\
(m^2_F)_{I\bar{J}}
 &\equiv& 
     \left(|m_{3/2}|^2 + \frac{\langle V_F \rangle}{M^2}\right) 
     \langle K_{I\bar{J}} \rangle 
\nonumber \\
&~& {}+ \langle F^{I'} \rangle \langle \bar{F}^{\bar{J}'} \rangle 
  \left \langle \partial_{I'} K_{I\bar{J}''} (K^{-1})^{\bar{J}''I''} 
        \partial_{\bar{J}'} K_{I''\bar{J}} 
      - \partial_{I'} \partial_{\bar{J}'}K_{I\bar{J}}
  \right \rangle \ ,
\label{mF}\\
(m^2_D)_{I\bar{J}}
 &\equiv&
     \langle D^\alpha \rangle 
     \left\langle \frac{\partial}{\partial \phi^I \bar{\phi}^{\bar J}}  
     \!\left(G_{I'} ( T^\alpha \phi)^{I'}\right) \right \rangle \ ,
\label{mD}
\end{eqnarray}
where
$m_{3/2}=\langle e^{K/2M^{2}} W/M^2 \rangle$ is the gravitino mass.
The magnitude of $m_{3/2}$ is expected to be $\order{1}$ TeV on the
phenomenological ground.
The first term in Eq.~(\ref{m}) originates from
the $F$-term scalar potential $V_F$ 
and so we will refer to it as the $F$-term scalar mass.
On the other hand, the second term, Eq.~(\ref{mD}), 
is the $D$-term contribution to scalar masses \cite{D-term,KMY}.
It is proportional to the charge of broken symmetry
and appears when the rank of gauge group lowers 
on the breakdown of gauge symmetry.

By taking the VEV of 
$(\partial V /\partial \phi^I)(T^\alpha \phi)^{I}$ and 
using the stationary condition,
we derive the useful formula for $\langle D^\alpha \rangle$,
\begin{eqnarray}
 & & \left[ (M_V^2)^{\alpha\beta} 
+ \left(\frac{\langle V_F \rangle}{M^2} + 2 |m_{3/2}|^2\right)
 \langle \mbox{Re} f_{\alpha\beta} \rangle \right]
\langle D^\beta \rangle 
\nonumber \\
 &~& 
  = \langle F^I \rangle \langle \bar{F}^{\bar{J}} \rangle 
    \left \langle 
    \frac{\partial}{\partial {\phi}^I \bar{\phi}^{\bar{J}}} 
    \!\left(G_{I'} ( T^\alpha \phi)^{I'}\right) 
    \right \rangle 
  + \frac{1}{2}\left \langle \frac{\partial}{\partial \phi^I} 
    \mbox{Re} f_{\beta \gamma}\right \rangle 
    \langle (T^\alpha \phi)^I \rangle 
    \langle D^\beta \rangle \langle D^\gamma \rangle \ ,
\label{<VI>}
\end{eqnarray}
where 
$(M_{V}^{2})^{\alpha \beta}=  
\left\langle  (\bar{\phi} T^{\beta})^{\bar{J}}
K_{I\bar{J}}  (T^\alpha \phi)^I\right\rangle$
is the mass matrix of the gauge bosons, up to the normalization 
factor including the gauge coupling constants.

We require that the SUSY is broken down 
by non-vanishing $F$-component VEVs of $\order{m_{3/2}M}$ and 
its effect is mediated through the gravitational interaction.
When the extra gauge boson mass is much larger than $m_{3/2}$,
the last term is negligibly small compared with other terms 
in Eq.~(\ref{<VI>}). Then the formula is simplified as
\begin{eqnarray}
\langle D^\beta \rangle
 = \langle F^I \rangle \langle \bar{F}^{\bar{J}} \rangle 
   \left \langle \frac{\partial}{\partial {\phi}^I 
   \bar{\phi}^{\bar{J}}} 
   \!\left(G_{I'} ( T^\alpha \phi)^{I'}\right)\right \rangle 
   \left({M}_V^{-2}\right)^{\alpha\beta} \ ,
\label{<Dbeta>}
\end{eqnarray}
where $({M}_V^{-2})^{\alpha\beta}$ is the inverse matrix of 
$({M}_V^{2})^{\alpha\beta}$.
The formula (\ref{<Dbeta>}) is the master equation 
in our analysis.

{}For a later convenience, we write down the formula of 
gaugino masses $M_{1/2}^\alpha$:
\begin{eqnarray}
M_{1/2}^\alpha \delta_{\alpha\beta} 
 = \frac{1}{2 \langle \mbox{Re} f_{\alpha\beta} \rangle}\,
   \langle F^I \rangle
   \left \langle 
   \frac{\partial}{\partial \phi^I} f_{\alpha\beta}
   \right \rangle \ .
\label{M1/2}
\end{eqnarray}

\section{Anomalous $U(1)$ $D$-term in heterotic string models}
\label{sec:het}

Effective SUGRA is derived from 4D string 
models taking
a field theory limit \cite{string}.
In this section, we review the $D$-term contribution for the 
anomalous $U(1)$ symmetry ($U(1)_A$)
in 4D heterotic string models \cite{KK}.
The K\"ahler potential $K(\phi^I, \bar{\phi}^{\bar{I}})$ and 
the gauge kinetic function 
$f_{\alpha \beta}(\phi^I)$ are given by
\begin{eqnarray}
K(\phi^I, \bar{\phi}^{\bar{I}})
 &=&{}- \ln (S + \bar{S} - 2 \delta_{GS}^A V_A)
      - \sum_a \ln(T^a + \bar{T}^a) 
\nonumber \\
 &~&{}+ \sum_{\kappa} \prod_a (T^a + \bar{T}^a)^{n_\kappa^a} 
        \bar{\phi}^{\bar \kappa}e^{2q_{\kappa}^{A}V_{A}}\phi^{\kappa}
      + \cdots \ ,
\label{K-het} \\
f_{\alpha \beta}(\phi^I)
 &=& k_{\alpha} S \delta_{\alpha\beta} 
      + \varepsilon_{\alpha}^a T^a \delta_{\alpha\beta} \ ,
\label{f-het} 
\end{eqnarray}
where $S$ is the dilaton field, $T^a$ are the moduli fields,
$\phi^\kappa$ are matter fields with modular weights  
$n_{\kappa}^a$ and $U(1)_A$ charges  $q_{\kappa}^A$,
and $V_A$ is the $U(1)_A$ vector superfield.\footnote{
In (\ref{K-het}) and (\ref{f-het}),
all fields stand for superfields with a same notation for 
chiral superfields and anti-chiral superfields as its scalar 
components.}
Also in the above, $k_{\alpha}$ is a Kac-Moody level (hereafter
we set $k_{\alpha}=1$, for simplicity), 
$\varepsilon_{\alpha}^a$ is a model-dependent
parameter coming from 1-loop correction and $\delta_{GS}^{A}$ is 
the GS coefficient of $U(1)_A$ given by
\begin{eqnarray}
\delta_{GS}^A = \frac{1}{192\pi^2} \sum_{\kappa} q_{\kappa}^A \ .
\label{GS} 
\end{eqnarray}
The $U(1)_A$ $D$-component is given by
\begin{eqnarray}
D^A
 = (\mbox{Re}f^{-1})_A 
   \left(\frac{\delta_{GS}^A}{S + \bar{S}}
 + \sum_{\kappa} \prod_a (T^a + \bar{T}^a)^{n_\kappa^a} 
   q_{\kappa}^A |\phi^\kappa|^2\right) \,,
\label{DA} 
\end{eqnarray}
where we neglect terms from higher order terms 
in $K(\phi^I, \bar{\phi}^{\bar{I}})$.
{}Following the custom in 4D SUGRA derived from string models,
we take the $M=1$ unit if no confusion is expected.

The $U(1)_A$ and its mixed anomalies due to matter fields 
are cancelled by the contribution from
the dilaton field which transforms non-linearly as $S \to S' = S 
+ i\delta_{GS}^A \theta(x)$ under $U(1)_A$.
Then the formula (\ref{<Dbeta>}) for $\langle D^A \rangle$ reads
\begin{eqnarray}
\langle D^A \rangle 
 = \frac{1}{({M}_V^{2})^{A}} 
   \left(
   2\delta_{GS}^A
   \frac{|\langle F^S \rangle |^2}{\langle S+\bar{S} \rangle^3}
 + \VEV{ \sum_{\kappa}\prod_a(T^a+\bar{T}^a)^{n_\kappa^a}
         q_\kappa^A\abs{F^\kappa}^2}
 + \cdots\ \right) \,.
\label{<DA>}
\end{eqnarray}
Here $({M}_V^{2})^{A}$ is given by
\begin{eqnarray}
({M}_V^{2})^{A} 
 = \frac{(\delta_{GS}^A)^2}{\langle S+\bar{S} \rangle^2}
 + \VEV{\sum_{\kappa}\prod_a(T^a+\bar{T}^a)^{n_\kappa^a}
        (q_\kappa^A)^2 \abs{\phi^\kappa}^2} \,,
\label{hatMV-A}
\end{eqnarray}
which can be rewritten 
with the help of the almost $D$-flatness condition of $U(1)_A$ into
\begin{eqnarray}
({M}_V^{2})^{A}
 = \frac{\delta_{GS}^A}{\langle S+\bar{S} \rangle}
   \left(\frac{\delta_{GS}^A}{\langle S+\bar{S} \rangle}
 - \frac{\VEV{ \sum_{\kappa}\prod_a (T^a+\bar{T}^a)^{n_\kappa^a}
               (q_\kappa^A)^2 \abs{\phi^\kappa}^2 }}%
        {\VEV{ \sum_{\kappa}\prod_a (T^a+\bar{T}^a)^{n_\kappa^a}
             \, q_\kappa^A \abs{\phi^\kappa}^2 }}
   \right) \,.
\label{hatMV-A2}
\end{eqnarray}
In explicit models, we find that 
$\delta_{GS}^A = \order{10^{-1}}$ -- $\order{10^{-2}}$.
Hence we will neglect terms with a higher order of $\delta_{GS}^A$.
With this assumption, 
the second term is dominant in Eq.~(\ref{hatMV-A}).

{}For simplicity, we treat the case with the overall moduli, 
i.e., $T = T^1 = T^2 = T^3$.
In this case, $\langle D^A \rangle$ is given by
\begin{eqnarray}
\langle D^A \rangle 
 &=& \frac{1}{(M_V^2)^A}
     \left(
     2\delta_{GS}^A
     \frac{|\langle F^S \rangle|^2}{\langle S+\bar{S} \rangle^3}
   + \VEV{ \sum_{\kappa}(T+\bar{T})^{n_\kappa}
           q_\kappa^A|F^\kappa|^2} \right.
\nonumber \\
 &~& \qquad\qquad
 {}+ \frac{|\langle F^T \rangle|^2}{\langle T+\bar{T} \rangle^2}
     \VEV{ \sum_{\kappa}(T+\bar{T})^{n_\kappa} 
           n_{\kappa}\left(n_{\kappa}  - 1\right) 
           q_\kappa^A \abs{\phi^\kappa}^2}
\nonumber \\
 &~& \qquad\qquad
 {}+ \left.
     \frac{\langle \bar{F}^{\bar{T}} \rangle}%
          {\langle T + \bar{T} \rangle} \,
     \VEV{ \sum_{\kappa}(T+\bar{T})^{n_\kappa} n_{\kappa} 
           q_\kappa^A F^\kappa \bar \phi^{\bar \kappa} }
   + \mbox{ h.c. }
     \right) \,.
\label{<DA>-T}
\end{eqnarray}

Here we consider the case that the dilaton and the overall 
moduli fields are dominant sources to the SUSY breaking, 
e.g. $\langle F^S \rangle$, $\langle F^T \rangle = \order{m_{3/2} M}$.
This situation is realized if $\VEV{\phi^\kappa} \ll \order{M}$ and 
$\VEV{\partial W /\partial \phi^\kappa} \ll \order{m_{3/2} M}$.
In this case, 
from the expression (\ref{D}) for $\langle F^I \rangle$,
we find that the $F$-terms $\langle F^S \rangle$ of
the chiral matter fields are induced as
\begin{equation}
\langle F^\kappa \rangle
 = \left( m_{3/2} - n_\kappa 
   \frac{\langle F^T \rangle}{\langle T + \bar T \rangle} 
   \right) \langle \phi^\kappa \rangle \ .
\end{equation}
Since the induced $\langle F^\kappa \rangle $ is much smaller 
than $\langle F^S \rangle$ and $\langle F^T \rangle$, 
the VEV $\langle V_F \rangle$ is simplified to
\begin{eqnarray}
\langle V_F \rangle 
 = \frac{|\langle F^S \rangle|^2}{\langle S+\bar{S} \rangle^2}
 + 3 \frac{|\langle F^T \rangle|^2}{\langle T+\bar{T} \rangle^2} 
 - 3 \abs{m_{3/2}}^2 \ .
\label{<VF>}
\end{eqnarray}
In Eq.~(\ref{<DA>-T}), however, 
the terms including $\langle F^\kappa \rangle$ 
are comparable with the other terms, and we obtain 
\begin{eqnarray}
\langle D^A \rangle 
 &=& \frac{1}{(M_V^2)^A}
     \left[
     \frac{\delta_{GS}^A}{S+\bar{S}} 
     \left(2 \frac{|\langle F^S\rangle|^2}{\langle S+\bar{S}\rangle^2}
   - \abs{m_{3/2}}^2\right) \right.
\nonumber \\
 &~& \qquad\quad
 {}- \left.\frac{|\langle F^T \rangle|^2}{\langle T+\bar{T} \rangle^2}
    \VEV{\sum_{\kappa}(T+\bar{T})^{n_\kappa} n_{\kappa} q_\kappa^A 
         \abs{\phi^\kappa}^2 }
    \right] \,.
\label{<DA>-T1}
\end{eqnarray}
By use of the parametrization
\begin{eqnarray}
\frac{\langle F^S \rangle}{\langle S+\bar{S} \rangle} 
\equiv \sqrt{3} 
C\abs{m_{3/2}} e^{i\alpha_S} \sin \theta \ ,  \qquad
\frac{\langle F^T \rangle}{\langle T+\bar{T} \rangle} 
\equiv C \abs{m_{3/2}} e^{i\alpha_T} \cos \theta \ ,
\label{FT:het}
\end{eqnarray}
the VEVs $\langle V_F \rangle$ and $\langle D^A \rangle$ 
can be expressed as
\begin{eqnarray}
\langle V_F \rangle
 &=& 3\left(C^2 - 1\right)\abs{m_{3/2}}^2 \ ,
\label{<VF>-para}\\
\langle D^A \rangle 
 &=& \abs{m_{3/2}}^2 
     \left[\left(1 - 6C^2 \sin^2\theta\right) 
 \frac{\VEV{ \sum_{\kappa}(T+\bar{T})^{n_\kappa}
       q_\kappa^A \abs{\phi^\kappa}^2 }}%
      {\VEV{ \sum_{\kappa}(T+\bar{T})^{n_\kappa} 
       (q_\kappa^A)^2 \abs{\phi^\kappa}^2 }} \right. 
\nonumber \\
 &~& \qquad\qquad\quad\,
 {}- \left. C^2 \cos^2\theta~~
 \frac{\VEV{ \sum_{\kappa}(T+\bar{T})^{n_\kappa} n_{\kappa} 
       q_\kappa^A \abs{\phi^\kappa}^2 }}%
      {\VEV{ \sum_{\kappa}(T+\bar{T})^{n_\kappa} 
       (q_\kappa^A)^2 \abs{\phi^\kappa}^2 }}\,
 \right] \,.
\label{<DA>-T2}
\end{eqnarray}
Here $C$ is a constant and $\theta$ is a parameter called the 
$\lq\lq$goldstino angle".

The gaugino masses $M_{1/2}^\alpha$ are calculated 
by use of Eq.~(\ref{M1/2}) to be
\begin{eqnarray}
M_{1/2}^\alpha  
= \frac{1}{2 \langle \mbox{Re} f_\alpha \rangle} 
\left(\langle F^S \rangle 
+ \varepsilon_\alpha \langle F^T  \rangle\right) \ .
\label{M1/2alpha:het}
\end{eqnarray}
To obtain gaugino masses of $\order{m_{3/2}}$, we need 
a dilaton dominant SUSY breaking
scenario in the weakly coupled region.
In the strongly coupled region, 
the moduli $F$-component can also lead to gaugino masses of 
$\order{m_{3/2}}$ \cite{Nilles:1997cm}.
In any case, $U(1)_A$ $D$-term contribution to scalar masses appears,
$(m^2_D)_I = q_I^A \langle D^A \rangle$,
and its magnitude is rather large as 
$\langle D^A \rangle = \order{m_{3/2}^2}$.
If $U(1)_A$ charges are different 
between the first and second families,
the non-universality among sfermion masses would be dangerous 
from the viewpoint of FCNC constraints.
On the other hand, with $U(1)_A$ $D$-term contribution 
we can relax the CCB and UFB bounds \cite{CCB}.

In a certain case \cite{SUSY-breaking:mech},  
the $F$-components of matter fields can also contribute 
to the breakdown of SUSY.\footnote{
See also Ref.~\cite{SUSY-breaking2}.
}
{}For instance,
$\langle F^\kappa \rangle$ contributes to the SUSY breaking
if $\VEV{\partial W /\partial \phi^\kappa} = \order{m_{3/2}M}$.
Then the dominant part of the $D^A$ condensation comes from the 
second term in the r.h.s. of Eq.~(\ref{<DA>-T}),
\begin{eqnarray}
\langle D^A \rangle 
 = \frac{\VEV{ \sum_{\kappa}(T+\bar{T})^{n_\kappa}
         q_\kappa^A \abs{F^\kappa}^2 }}%
        {\VEV{ \sum_{\kappa}(T+\bar{T})^{n_\kappa} 
         (q_\kappa^A)^2 \abs{\phi^\kappa}^2 }} \ .
\label{<DA>-T-matter}
\end{eqnarray}
The magnitude is estimated \cite{K} as 
$\langle D^A \rangle = \order{m_{3/2}^2/\delta_{GS}^A}$.

\section{Anomalous $U(1)$ $D$-terms in Type~I Models}
\label{sec:typeI}

Next we turn to the type~I case.
In general, a 4D type~I model has more than one anomalous $U(1)$
symmetries, i.e. $\prod_i U(1)_i$.
We denote the $U(1)_i$ vector multiplet by $V_i$.
The K\"ahler potential $K(\phi^I, \bar{\phi}^{\bar{I}})$ 
is given by\footnote{
See for effective low-energy Lagrangian of type~I models 
Ref.~\cite{IIB} and references therein.}
\begin{eqnarray}
K(\phi^I, \bar{\phi}^{\bar{I}})
 &=& \hat{K}\!\left(M_\ell + \bar{M_\ell} 
                  - 2 \sum_i (\delta_{GS})_i^\ell V_i\right)
  - \ln(S + \bar{S})
  - \sum_{a} \ln(T^a + \bar{T}^a) 
\nonumber \\
 &~& 
{}+ \sum_{\kappa}\prod_a 
    (S + \bar{S})^{n_\kappa^s}(T^a + \bar{T}^a)^{n_\kappa^a} 
    \bar{\phi}^{\bar \kappa} e^{2q_{\kappa}^{i}V_{i}}\phi^{\kappa}
  + \cdots \ ,
\label{K-2B} 
\end{eqnarray}
where 
chiral matter fields $\phi^\kappa$ have the ``modular weights'' 
$n_{\kappa}^s$ and $n_{\kappa}^a$ with respect of $S$ and $T^a$, 
and $(\delta_{GS})_i^\ell$ are model-dependent GS coefficients.
Here, $M_\ell$ is a twisted moduli field 
associated with the $\ell$-th fixed point.
{}For simplicity, we use the notation $m_\ell$ defined by 
$m_\ell \equiv 
M_\ell + \bar{M_\ell} - 2 \sum_i (\delta_{GS})_i^\ell V_i$ hereafter.
The complete form of $\hat{K}$ is unknown, 
but in the orbifold limit $M_\ell \rightarrow 0$, 
it takes the tree level form \cite{Mk}
\begin{eqnarray}
\hat{K}(m_\ell) = \frac{1}{2}\, m_\ell^2 \ .
\label{hatK} 
\end{eqnarray}
The $M_\ell$-dependence of K\"ahler metric of $\phi^\kappa$ 
is also unclear. 
In the orbifold limit,
the K\"ahler metric $K_{\kappa \bar \kappa}$ 
does not depend on the twisted moduli $M_\ell$ as in Eq.~(\ref{K-2B}).
{}For a large value of $M_\ell$, however, it would receive 
a correction $\Delta K_{\kappa \bar \kappa}(M,\bar M)$.

The gauge kinetic function $f_{\alpha \beta}(\phi^I)$ is given by
\begin{eqnarray}
f_{\alpha \beta}(\phi^I)
 = \hat{f}(S, T^a) \delta_{\alpha\beta} 
 + \sum_\ell s^\alpha_\ell M_\ell \delta_{\alpha\beta} \ ,
\label{f-2B} 
\end{eqnarray}
where $s_{\ell}^\alpha$ is a model-dependent constant.
The first term is D-brane dependent, e.g.,
$\hat{f}(S, T^a) = S$ for gauge groups from D9-branes and 
$\hat{f}(S, T^a) = T^a$ for gauge groups from D5$_a$-branes.
The $U(1)_i$ and mixed anomalies due to matter fields are cancelled 
by the contribution from the twisted moduli fields which transform as 
$M_\ell \to M'_\ell 
= M_\ell + i(\delta_{GS})_i^\ell \theta(x)$ under $U(1)_i$.

The $U(1)_i$ $D$-components are given by
\begin{eqnarray}
D^i
 = (\mbox{Re}f^{-1})_i 
   \left({} -(\delta_{GS})_i^\ell\,
   \frac{\partial\hat{K}}{\partial m_\ell} 
 + \sum_{\kappa} \prod_a (S+\bar{S})^{n_\kappa^s}
   (T^a + \bar{T}^a)^{n_\kappa^a} q_{\kappa}^i |\phi^\kappa|^2
   \right) \,,
\label{Di} 
\end{eqnarray}
where we assume that $U(1)$ kinetic mixing is absent for simplicity.
According to the formula (\ref{<Dbeta>}) for 
$D$-term condensation, we obtain
\begin{eqnarray}
\langle D^i \rangle
 &=& \frac{1}{({M}_V^{2})^{i}} 
     \left({}- (\delta_{GS})_i^\ell 
     \left\langle 
     \frac{\partial^3 \hat{K}}%
          {\partial m_\ell \partial m_{\ell'} \partial m_{\ell''}}
     F^{M_{\ell'}} \bar{F}^{\bar{M}_{\bar{\ell}''}}
     \right \rangle \right.
\nonumber\\
 &~& \qquad\qquad\!\!
 {}+ \left.
     \VEV{ \sum_{\kappa}\prod_a
           (S+\bar{S})^{n_\kappa^s}(T^a+\bar{T}^a)^{n_\kappa^a}
           q_\kappa^i |F^\kappa|^2 }
   + \cdots\ \right) \,.
\label{<Di>}
\end{eqnarray}
Here $({M}_V^{2})^{i}$ is (essentially) 
the $U(1)_i$ gauge boson mass given by
\begin{eqnarray}
({M}_V^{2})^{i}
 &\equiv& 
     (\delta_{GS})_i^\ell (\delta_{GS})_i^{\ell'}
     \left \langle 
     \frac{\partial^2 \hat{K}}{\partial m_\ell \partial m_{\ell'}} 
     \right\rangle 
   + \VEV{ \sum_{\kappa}\prod_a(S+\bar{S})^{n_\kappa^s}
           (T^a+\bar{T}^a)^{n_\kappa^a}
           (q_\kappa^i)^2|\phi^\kappa|^2 }
\nonumber \\
 &=& (\delta_{GS})_i^\ell  (\delta_{GS})_i^{\ell'} 
     \left \langle 
     \frac{\partial^2 \hat{K}}{\partial m_\ell \partial m_{\ell'}}
     \right\rangle
\nonumber \\
 & &
 {}+ (\delta_{GS})_i^\ell 
     \left\langle \frac{\partial \hat{K}}{\partial m_\ell}
     \right\rangle 
     \frac{%
     \VEV{ \sum_{\kappa}\prod_a
           (S+\bar{S})^{n_\kappa^s} (T^a+\bar{T}^a)^{n_\kappa^a}
           (q_\kappa^i)^2\abs{\phi^\kappa}^2 }}{%
     \VEV{ \sum_{\kappa}\prod_a(S+\bar{S})^{n_\kappa^s}
           (T^a+\bar{T}^a)^{n_\kappa^a}q_\kappa^i
           \abs{\phi^\kappa}^2 }} \ .
\label{hatMV-i}
\end{eqnarray}
where we have used the almost $D$-flatness conditions of $U(1)_i$.
In the case with the canonical K\"ahler potential (\ref{hatK}), 
the $({M}_V^{2})^{i}$ is reduced to
\begin{eqnarray}
({M}_V^{2})^{i} \equiv 
   \left( (\delta_{GS})_i^\ell\right)^2
 + (\delta_{GS})_i^\ell \langle m_\ell \rangle \,
   \frac{\VEV{ \sum_{\kappa}\prod_a(S+\bar{S})^{n_\kappa^s}
         (T^a+\bar{T}^a)^{n_\kappa^a}(q_\kappa^i)^2
         \abs{\phi^\kappa}^2 }}%
        {\VEV{ \sum_{\kappa}\prod_a(S+\bar{S})^{n_\kappa^s}
         (T^a+\bar{T}^a)^{n_\kappa^a}q_\kappa^i
         \abs{\phi^\kappa}^2 }} \ .
\label{hatMV-i2}
\end{eqnarray}
If $\langle m_\ell \rangle \ll \order{\delta_{GS} }$, 
the first term is dominant in Eq.~(\ref{hatMV-i2}),
unlike the heterotic case (\ref{hatMV-A}).

Again we treat the case with the overall moduli, 
i.e., $T = T^1 = T^2 = T^3$,
and denote $n_\kappa = \sum_a n_\kappa^a$.
Then the $\langle D^i \rangle$ in Eq.~(\ref{<Di>})
can explicitly be written down as
\begin{eqnarray}
\langle D^i \rangle 
 &\!\!=\!\!& 
     \frac{1}{({M}_V^{2})^{i}} 
     \left(
 {}- (\delta_{GS})_i^\ell 
     \left\langle 
     \frac{\partial^3 \hat{K}}%
          {\partial m_\ell \partial m_{\ell'} \partial m_{\ell''}}
     F^{M_{\ell'}} \bar{F}^{\bar{M}_{\bar{\ell}''}} 
     \right\rangle \right. 
\nonumber \\
 &~& \qquad\qquad\!
 {}+ \VEV{ \sum_{\kappa}
           (S+\bar{S})^{n_\kappa^s} (T+\bar{T})^{n_\kappa} 
           q_\kappa^i|F^\kappa|^2 }
\nonumber \\
 &~& \qquad\qquad\!
 {}+ \frac{|\langle F^S \rangle|^2}{\langle S +\bar{S} \rangle^2}
     \VEV{ \sum_{\kappa}(S+\bar{S})^{n_\kappa^s}(T+\bar{T})^{n_\kappa} 
           n_{\kappa}^s (n_{\kappa}^s - 1) q_\kappa^i
           \abs{\phi^\kappa}^2 }
\nonumber \\
 &~& \qquad\qquad\!
 {}+ \frac{|\langle F^T \rangle|^2}{\langle T +\bar{T} \rangle^2}
     \VEV{ \sum_{\kappa}(S+\bar{S})^{n_\kappa^s}(T+\bar{T})^{n_\kappa} 
           n_{\kappa}(n_{\kappa} - 1) q_\kappa^i 
           \abs{\phi^\kappa}^2 }
\nonumber \\
 &~& \qquad\qquad\!
 {}+ \frac{|\langle F^S \rangle|}{\langle S+\bar{S} \rangle}
     \frac{|\langle \bar{F}^{\bar{T}} \rangle|}%
          {\langle T+\bar{T} \rangle}
     \VEV{ \sum_{\kappa}(S+\bar{S})^{n_\kappa^s}(T+\bar{T})^{n_\kappa} 
           n_{\kappa}^s n_{\kappa} q_\kappa^i 
           \abs{\phi^\kappa}^2 }
   + \mbox{h.c.}
\nonumber \\
 &~& \qquad\qquad\!
 {}+ \frac{\langle \bar{F}^{\bar{S}} \rangle}%
          {\langle S + \bar{S} \rangle} 
     \VEV{ \sum_{\kappa}(S+\bar{S})^{n_\kappa^s}(T+\bar{T})^{n_\kappa} 
           n_{\kappa}^s q_\kappa^i 
           F^\kappa \bar{\phi}^{\bar{\kappa}} }
   + \mbox{h.c.}
\nonumber \\
 &~& \qquad\qquad\!
 {}+ \left.
     \frac{\langle \bar{F}^{\bar{T}} \rangle}%
          {\langle T + \bar{T} \rangle} 
     \VEV{ \sum_{\kappa}(S+\bar{S})^{n_\kappa^s}(T+\bar{T})^{n_\kappa} 
           n_{\kappa} q_\kappa^i 
           F^\kappa \bar{\phi}^{\bar{\kappa}} }
   + \mbox{h.c.}\ 
     \right) \,.
\label{<Di>-T}
\end{eqnarray}

In the following,
we mainly consider the case that 
SUSY is broken by the dilaton, the overall moduli 
and/or the twisted moduli fields,
$\langle F^S \rangle$, $\langle F^T \rangle$, 
$\langle F^{M_\ell} \rangle = \order{m_{3/2} M}$.
This situation is realized if 
$\VEV{\phi^\kappa} \ll \order{M}$ and 
$\VEV{\partial W/\partial \phi^\kappa} \ll \order{m_{3/2}M}$.
In this case, since
$\VEV{F^\kappa} = \order{m_{3/2}\VEV{\phi^\kappa}}$,
the VEV $\VEV{V_F}$ is simplified as
\begin{eqnarray}
\langle V_F \rangle 
= \frac{|\langle F^S \rangle|^2}{\langle S+\bar{S} \rangle^2}
 + 3 \frac{|\langle F^T \rangle|^2}{\langle T+\bar{T} \rangle^2} 
+ \sum_{\ell, \ell'} \left\langle 
\frac{\partial^2 \hat{K}}{\partial m_\ell 
\partial m_{\ell'}}
{}F^{M_{\ell}} \bar{F}^{\bar{M}_{\bar{\ell}'}} \right\rangle
- 3 \abs{m_{3/2}}^2 \ .
\label{<VFIIB>}
\end{eqnarray}
To calculate the $\langle D^i \rangle$, however, it is important to
note that $\langle F^\kappa \rangle$ is induced as
\begin{equation}
\langle F^\kappa \rangle
 = \left( m_{3/2} 
 - n^s_\kappa \frac{\langle F^S \rangle}{\langle S + \bar S \rangle}
 - n_\kappa   \frac{\langle F^T \rangle}{\langle T + \bar T \rangle} 
   \right)
   \langle \phi^\kappa \rangle \ .
\end{equation}
Then a careful calculation leads to
\begin{eqnarray}
\langle D^i \rangle 
 &=& \frac{1}{({M}_V^{2})^{i}} 
    \left(
    {} - (\delta_{GS})_i^\ell 
    \left\langle 
    \frac{\partial^3 \hat{K}}%
         {\partial m_\ell \partial m_{\ell'} \partial m_{\ell''}}
    F^{M_{\ell'}} \bar{F}^{\bar{M}_{\bar{\ell}''}} 
    \right\rangle \right. 
\nonumber \\
 &~& \qquad\qquad\!
    {}+ \abs{m_{3/2}}^2 
    \VEV{ \sum_{\kappa}
          (S+\bar{S})^{n_\kappa^s} (T+\bar{T})^{n_\kappa} 
          q_\kappa^i \abs{\phi^\kappa}^2 }
\nonumber\\
 &~& \qquad\qquad\!
    {}- \frac{|\langle F^S \rangle|^2}{\langle S+\bar{S} \rangle^2}
    \VEV{ \sum_{\kappa}(S+\bar{S})^{n_\kappa^s}
          (T+\bar{T})^{n_\kappa} n_{\kappa}^s
          q_\kappa^i \abs{\phi^\kappa}^2 }
\nonumber\\
 &~& \qquad\qquad\!
    {}- \left. 
    \frac{|\langle F^T \rangle|^2}{\langle T+\bar{T} \rangle^2}
    \VEV{ \sum_{\kappa}(S+\bar{S})^{n_\kappa^s}(T+\bar{T})^{n_\kappa} 
          n_{\kappa} q_\kappa^i \abs{\phi^\kappa}^2 }
    \right) \,.
\label{<Di>-T1}
\end{eqnarray}
The expressions (\ref{<Di>-T}) or (\ref{<Di>-T1}) are our main results
for the $D$-term condensation in type~I models. 
Note that $\langle D^i\rangle$ becomes independent of 
$\langle F^{M_\ell} \rangle$ if the third derivative of $\hat{K}$ 
vanishes;
$\VEV{\partial^3 \hat{K}/\partial m_\ell \partial m_{\ell'} 
\partial m_{\ell''}} \ll \order{(\delta_{GS})_i^\ell}$.

The soft terms can be calculated 
by using the parametrization similar to Eq.~(\ref{FT:het}),  
\begin{eqnarray}
\frac{\langle F^S \rangle}{\langle S+\bar{S} \rangle} 
 &\equiv& 
  \sqrt{3} C \abs{m_{3/2}} e^{i\alpha_S} \sin \theta \ , \qquad
\frac{\langle F^T \rangle}{\langle T+\bar{T} \rangle} 
 \equiv C \abs{m_{3/2}}e^{i\alpha_T} \cos \theta \sin \phi \ ,
\nonumber \\
\langle F^{M_\ell} \rangle
 &\equiv&
  \sqrt{3} C \abs{m_{3/2}} e^{i\alpha_\ell}
  \Phi_\ell \cos \theta \cos \phi \ , \qquad
\sum_\ell \Phi_\ell^2 = 1 \ ,
\label{FT}
\end{eqnarray}
where we assumed that the K\"ahler metric of the twisted moduli is
diagonal, and $\theta$, $\phi$ and $\Phi_\ell$ 
are 
$\lq\lq$goldstino angles".
Then $\langle V_F \rangle$ is the same as in Eq.~(\ref{<VF>-para}),
and the $\langle D^i \rangle$ becomes
\begin{eqnarray}
\langle D^i \rangle 
 &\!=\!& 
     \frac{\abs{m_{3/2}}^2}{({M}_V^{2})^{i}}
     \left(
 {}- 3 C^2 (\delta_{GS})_i^\ell 
     \left \langle
     \frac{\partial^3 \hat{K}}%
          {\partial m_\ell \partial m_{\ell'} \partial m_{\ell''}} 
     \right\rangle
     \Phi_{\ell'} \Phi_{\ell''} \cos^2 \theta \cos^2 \phi 
  + (\delta_{GS})_i^\ell 
     \left \langle 
     \frac{\partial \hat{K}}{\partial m_\ell}
     \right\rangle
     \right.
\nonumber \\
 &~& \qquad\qquad
 {}- 3 C^2 \sin^2 \theta 
    \VEV{ \sum_{\kappa}
          (S+\bar{S})^{n_\kappa^s}(T+\bar{T})^{n_\kappa} 
          n_{\kappa}^s q_\kappa^i \abs{\phi^\kappa}^2 }
\nonumber \\
 &~& \qquad\qquad
 {}- \left. C^2 \cos^2 \theta \sin^2 \phi 
    \VEV{ \sum_{\kappa}
          (S+\bar{S})^{n_\kappa^s}(T+\bar{T})^{n_\kappa}
          n_{\kappa} q_\kappa^i \abs{\phi^\kappa}^2 } \ 
    \right) \,.
\label{<Di>-T2}
\end{eqnarray}

The gaugino masses $M_{1/2}^\alpha$ are calculated 
by use of Eqs.~(\ref{M1/2}) and (\ref{f-2B}),
\begin{eqnarray}
M_{1/2}^\alpha  
 &=& \frac{1}{2 \langle \mbox{Re} f_\alpha \rangle} 
     \left(\langle F^S \rangle 
   + \sum_\ell s_\ell^{\alpha} \langle F^{M_\ell} \rangle\right)
     \qquad \mbox{for D9-branes} \ ,
\nonumber \\
M_{1/2}^\alpha  
 &=& \frac{1}{2 \langle \mbox{Re} f_\alpha \rangle} 
     \left(\langle F^{T} \rangle 
   + \sum_\ell s_\ell^{\alpha} \langle F^{M_\ell} \rangle\right) 
     \qquad \mbox{for D5$_a$-branes} \ .
\label{M1/2alpha}
\end{eqnarray}
To obtain sizable gaugino masses of $\order{m_{3/2}}$, we need 
the dilaton and/or twisted moduli
dominant SUSY breaking
scenario on D9-branes, and the overall moduli and/or twisted moduli
dominant SUSY breaking
scenario on D5$_a$-branes.
If the dilaton and/or an overall moduli dominant SUSY breaking occur,
the magnitude of $U(1)_i$ $D$-term can be sizable as 
$\langle D^i \rangle = \order{m_{3/2}^2}$.
On the other hand, the magnitude of $\langle D^i \rangle$ can be
small if the twisted moduli fields dominate the SUSY breaking; in this case, 
$\langle D^i \rangle$ is negligibly small if 
$\langle \partial^3 \hat{K}/\partial m_\ell \partial m_{\ell'} 
\partial m_{\ell''} \rangle \ll \order{(\delta_{GS})_i^\ell}$
and 
$\langle m_\ell \rangle \ll \order{(\delta_{GS})_i^\ell}$.

Up to here, we have assumed that SUSY is broken 
by the $F$-terms of $S$, $T$ or $M_\ell$. 
Alternatively we can suppose, as in the heterotic case, 
that there exists dynamical superpotential $W$ of 
chiral matter fields $\phi^\kappa$ so that 
$\langle \partial W /\partial \phi^\kappa \rangle = \order{m_{3/2} M}$.
In such case, the dominant part of the $D^i$ condensation 
comes from the second term in the r.h.s. of Eq.~(\ref{<Di>-T}),
\begin{eqnarray}
\langle D^i \rangle 
 = \frac{\VEV{ \sum_{\kappa}
         (S+\bar{S})^{n_\kappa^s}(T+\bar{T})^{n_\kappa} 
         q_\kappa^i |F^\kappa|^2 }}%
        {(M_{V}^2)^i} \ .
\label{<Di>-T-matter}
\end{eqnarray}
The magnitude is estimated as 
$\langle D^i \rangle = \order{m_{3/2}^2M^2/( M_{V}^2)^i}$.

\section{Phenomenological implications}
\label{sec:implications}

In this section, we discuss phenomenological implications 
of $D$-term contributions.
An important point is that 
the FI terms depend on the twisted moduli fields in type~I models, 
while such role is played by the dilaton field in heterotic models. 
Here let us compare our result (\ref{<Di>-T1}) in type~I models 
with the $D$-term (\ref{<DA>-T1}) in heterotic models.

The first term of the $D$-term condensation (\ref{<Di>-T1}) 
is negligibly small, 
when the canonical term is dominant in $\hat K(M_\ell,\bar M_\ell)$.
As a result, the $D$-term condensation 
does not depend on $F^{M_\ell}$ explicitly. 
Recall that 
the $D$-term condensation (\ref{<DA>-T1}) in heterotic models 
depends explicitly on $F^S$.
The difference is originated from the different forms of 
K\"ahler potential between $M_\ell$ and $S$.
If $\hat K(M_\ell,\bar M_\ell)$ is the logarithmic form like $S$, 
this difference would disappear.

{}For the remaining terms in Eq.~(\ref{<Di>-T1}), 
we can estimate the order of magnitudes by using the fact that 
the $D$-term (\ref{Di}) almost vanishes.
The second term 
is proportional to the FI terms as in the heterotic case.
The last two terms can be estimated as  
\begin{eqnarray}
\VEV{ \sum_{\kappa}(S+\bar{S})^{n_\kappa^s}(T+\bar{T})^{n_\kappa} 
      n_{\kappa}^s q_\kappa^i \abs{\phi^\kappa}^2 }
 &=& \order{ (\delta_{GS})_i^\ell 
     \left \langle 
     \frac{\partial \hat{K}}{\partial m_\ell}
     \right\rangle } \,, 
\\
\VEV{ \sum_{\kappa}(S+\bar{S})^{n_\kappa^s}(T+\bar{T})^{n_\kappa} 
      n_{\kappa} q_\kappa^i \abs{\phi^\kappa}^2 }
 &=& \order{ (\delta_{GS})_i^\ell 
     \left \langle 
     \frac{\partial \hat{K}}{\partial m_\ell}
     \right\rangle } \,. 
\end{eqnarray}
Therefore, we have 
\begin{eqnarray}
\langle D^i \rangle
 = m^2_{3/2} \times 
   \order{ \frac{1}{(M_V^2)^i} (\delta_{GS})_i^\ell
   \left \langle 
   \frac{\partial \hat{K}}{\partial m_\ell}
   \right\rangle } \,,
\end{eqnarray}
and its magnitude depends on 
$\langle m_\ell \rangle /(\delta_{GS})_i^\ell$
as is seen from Eq.~(\ref{hatMV-i2}).
Notice that unlike the dilaton VEV in the heterotic case,
the VEVs of twisted moduli fields $m_\ell$ 
can be taken as arbitrary value, depending on stabilization 
mechanism \cite{Abel:2000tf,Ciesielski:2002fs,Higaki:2003zk}.

If $\order{\langle m_{\ell} \rangle/(\delta_{GS})_i^\ell} \geq 1$,
$D$-term condensations are sizable 
and their order is $\order{m^2_{3/2}}$.
They significantly affect s-spectra. 
This situation is the same as that in the heterotic case.
{}For example, CCB and UFB directions have been studied 
for type~I models in Ref.~\cite{Abel:2000bj} 
with the string scale varied.
Hence, $D$-term contributions have important effects like 
the heterotic case if their magnitudes are $\order{m_{3/2}^{2}}$.

On the other hand, it is possible that the VEVs of 
twisted moduli fields $m_\ell$ are suppressed, i.e.,
$O\!\left(\langle m_{\ell} \rangle/(\delta_{GS})_i^\ell\right) \ll{}1$. 
In this case, $D$-term contribution can be suppressed.
This is a sharp contrast to the heterotic case 
where $D$-term contribution cannot be suppressed
without fine-tuning. 

To be concrete, 
let us first discuss the dilaton-dominant SUSY breaking
with $\langle V_F \rangle = 0$ in type~I models.
{}For comparison with the heterotic models,
we consider the case that 
the gauge multiplets originate from D$9$ branes and 
chiral matter fields originate from open strings, one of 
whose end is on the D$9$ branes.
In this case, the gaugino mass is obtained 
\begin{equation}
M_{1/2} = \sqrt{3} m_{3/2} \ ,
\end{equation}
where we have taken $\abs{s_{\ell}^{\alpha}\VEV{m_{\ell}}} \ll$ Re$S$.
Since $n_\kappa^s = 0$, the $F$-term scalar masses 
are universal, i.e.,
\begin{equation}
(m^2_F)_I = \abs{m_{3/2}}^2 \ .
\end{equation}
This spectrum is the same as the dilaton-dominant 
SUSY breaking in heterotic models.
In addition, we have to add the $D$-term contribution, 
$(m^2_D)_I = q^i_I \VEV{D^i}$ 
with $\VEV{D^i}$ given by Eq.~(\ref{<Di>-T2}).
When $\abs{\VEV{m_\ell}/(\delta_{GS})_i^\ell} \ll 1$, 
the $D$-term contribution becomes simplified to
\begin{equation}
(m^2_D)_I
 = q^i_I \abs{m_{3/2}}^2 
   \frac{\VEV{m_\ell}}{(\delta_{GS})_i^\ell} \ .
\label{mD:typeI}
\end{equation}
Thus, if $\abs{\VEV{m_\ell}/(\delta_{GS})_i^\ell} \ll 1$, 
the $D$-term contribution is small 
and the total soft scalar masses become almost universal.
This has important implications on FCNC constraints 
as well as CCB and UFB bounds.
That is, if this $U(1)$ symmetry is relevant to 
the flavor symmetry, the suppressed $D$-term contribution 
would be favorable to avoid  FCNC constraints.\footnote{
See e.g. Ref.~\cite{King:2003kf}, where 
such flavor $U(1)$ symmetry is discussed in a type-I inspired model.
}
{}For this purpose, we need to realize a suppression like 
$\abs{\VEV{m_\ell}/(\delta_{GS})_i^\ell} \leq \order{10^{-2}}$ 
for $m_{3/2}=\order{100}$ GeV.
Whether it is possible or not,
depends on the stabilization mechanism of $M_\ell$.

Next let us consider the case that 
the single twisted moduli field $M_\ell$ gives 
a dominant source in the SUSY breaking.
In this case, the gaugino mass is written as 
\begin{eqnarray}
M_{1/2}^{\alpha} 
 = \frac{\sqrt{3}}{2}\,s_{\ell}^{\alpha} g_{\alpha}^{2} m_{3/2} \ .
\end{eqnarray}
It is interesting to note that
if $s_{\ell}^{\alpha}$ are proportional to 
the coefficients of 1-loop beta function of gauge couplings,
like the case of `mirage gauge coupling unification'
in Ref.~\cite{Ibanez:1999st}, 
this spectrum of gaugino masses resembles
that in the anomaly mediation scenario \cite{Randall:1998uk}.
Since K\"ahler potential of matter fields 
does not depend on $M_\ell$ for small $\langle M_\ell \rangle$, 
the $F$-term scalar masses 
are universal, i.e.,
\begin{equation}
(m^2_F)_I = \abs{m_{3/2}}^2 \ .
\end{equation}
When $\abs{\VEV{m_\ell}/(\delta_{GS})_i^\ell} \ll 1$, 
the $D$-term contribution becomes
\begin{equation}
(m^2_D)_I
 = q^i_I \abs{m_{3/2}}^2 
   \left(
{}-\frac{3 C^2}{(\delta_{GS})_i^\ell} 
   \left\langle 
   \frac{\partial^3 \hat{K}}%
         {\partial m_\ell \partial m_{\ell'} \partial m_{\ell''}} 
    \right\rangle \Phi_{\ell'} \Phi_{\ell''} 
 + \frac{\VEV{m_\ell}}{(\delta_{GS})_i^\ell} 
   \right) \,.
\end{equation}
Thus, if $\VEV{\partial^3 \hat{K}/%
\partial m_\ell \partial m_{\ell'} \partial m_{\ell''}} %
\ll \order{(\delta_{GS})_i^\ell}$,
the $D$-term contribution is small 
and the total soft scalar masses become almost universal.

We note that
even if flavor-dependent $D$-term contributions can be suppressed, 
radiative corrections due to the gaugino mass of 
(gauged) flavor $U(1)$ symmetry would be important.
That might generate sizable effects on FCNC 
processes \cite{Kobayashi:2002mx}
when the gauged $U(1)$ is relevant to the flavor symmetry.
\section{Conclusion}
\label{sec:conclusion}


We have studied the $D$-term contribution 
for anomalous $U(1)$ symmetries in type~I models.
Specifically we have derived general formula
for the $D$-term contribution,
assuming that the dominant source of SUSY breaking
is given by $F$-terms of the dilaton, (overall) moduli or
twisted moduli fields.

We have also observed that
there are several differences in the $D$-term contributions 
between the heterotic and type~I models.
One of the important differences is that 
the $D$-term contribution in type~I models 
depends on the VEVs of twisted moduli fields, 
while that in heterotic models depends on the dilaton VEV.
The former can be taken as arbitrary values, 
although it depends on the stabilization mechanism.
That is, whether $D$-term contributions are sizable 
or can be suppressed, depends on the VEVs of 
twisted moduli.
This aspect will be important for 
CCB/UFB bounds and FCNC constraints.

The observed differences 
originate from the fact that in type~I models,
K\"ahler potential $\hat{K}$ of twisted moduli fields $M_\ell$
can take different forms from the dilaton K\"ahler potential.
{}For instance, 
if the $\hat{K}$ takes the tree level form (\ref{hatK}),
the $M_\ell$-dependent FI term vanishes
in the limit $\VEV{M_\ell}\rightarrow0$, 
and consequently, the anomalous $U(1)$ $D$-term contribution 
also vanishes in that limit.
Our results, e.g., Eq.~(\ref{mD:typeI}),
are consistent with this property.
Another remark concerns the sign of the FI term;
The FI term in type~I models can take both signs
depending on the sign of the twisted moduli VEVs.
Again this property is sharp contrast to the heterotic case.
{}Further phenomenological impact of these properties 
will be discussed elsewhere.

\section*{Acknowledgment}

The authors thank the Yukawa Institute for Theoretical Physics
at Kyoto University, where a part of this work was done
during the YITP-W-03-04 on ``Progress in Particle Physics''.
Y.~K.\/ is supported in part by the Grant-in-Aid for 
Scientific Research from Ministry of Education, Science, 
Sports and Culture of Japan (\#13135217) and (\#15340078).
T.~K.\/ is supported in part by the Grant-in-Aid for 
Scientific Research from Ministry of Education, Science, 
Sports and Culture of Japan (\#14540256).


\end{document}